# Depth to magnetic source estimation using TDX contour

*Hammed A. Oyekan\* , Department of Geosciences, King Fahd University of Petroleum and Minerals (KFUPM)*

**Summary**

Accurate depth estimation of magnetic sources plays a crucial role in various geophysical applications, including mineral exploration, resource assessments, regional hydrocarbon exploration, and geological mapping. Thus, this abstract presents a fast and simple method of estimating the depth of a magnetic body using the TDX derivative of the total magnetic field. TDX is a first-order derivative of the magnetic field that, in addition to edge detection, is less affected by noise, allowing for better depth resolution. The reduced sensitivity to noise enables a clearer estimation of depth and enhances the accuracy of the depth determination process. The TDX, as a variant of the phase derivative, is independent of magnetization and can be used to identify the edge of a magnetic body. In addition to excelling at edge detection, they can also estimate the depth of the magnetic source producing the anomalies. In this study, we explore the utilization of contour of the TDX derivative for estimating depth, assuming a vertical contact source. We demonstrate the effectiveness of the method using a two-prism block model and a simple bishop model with a uniform susceptibility of 0.001 cgs. The results agree with the known depth, providing evidence of the reliability of the method despite the restrictive nature of the assumption, especially for the Bishop model, where there are numerous fault structures.

**Introduction**

Recently, it has become increasingly important to develop fast and reliable techniques for estimating the subsurface position of the magnetic source causing magnetic anomalies. The need for an automated method arises from the need to qualitatively and quantitatively interpret the large volume of magnetic data being collected nowadays for both exploration and environmental purposes (Ahmed Salem et al. 2008). Researchers have developed various techniques to achieve these objectives for both profile and grid depth estimation. The evolution of these techniques can be traced back to the period before the introduction of Euler deconvolution in magnetic interpretation, in what could be referred to as the "pre-Euler era" of interpretation, during which graphical methods and several curve matching techniques such as Peters method (Peters 1949), horizontal slope distance (HSD) of (Vacquier et al. 1951), half-width method (Nettleton 1940), and several semi-automatic profile-based methods like Naudy (Naudy 1971) and Wenner method were introduced.

Subsequently, the "Euler era" ensued, which ushered in more advanced methods of depth interpretation techniques, started with the introduction of conventional 2D (Thompson 1982) and 3D Euler (A. B. Reid et al. 1990) deconvolution techniques. Faced with many limitations, especially being computationally intensive with too many depth solutions to contend with, the Euler techniques metamorphose as a result of a multitude of subsequent developments and enhancements to the approach. These include the Extended 2D Euler, Constrained 2D Euler, Laplacian XY Euler, Tensor, Tilt, Hilbert Euler methods, and several others. This era had a significant and revolutionary impact on the interpretation of magnetic data, facilitating the rapid calculation of the depth of a magnetic body. In the current period, the post-Euler era, researchers have developed more sophisticated and reliable techniques for analyzing the spectral characteristics of individual anomalies. These methods include the utilization of analytical signal, local wavenumber, tilt, and the power spectrum method. These advancements have facilitated a systematic examination of the spectral content associated with anomalies.

Even with the advancement in techniques, interpretation difficulties still exist with magnetic anomalies, as they are characterized by both positive and negative components. Mathematical derivatives have been used to resolve the ambiguities in interpreting the anomalies after the necessary mathematical transformation, which assumes vertical magnetization, has been applied. Amplitude derivatives have been used to locate the edges of magnetic bodies. However, they are sensitive to the magnitude of magnetization, making it difficult to identify smaller anomalies of interest and confidently estimate their depth in the presence of larger magnetic anomalies. Phase derivatives and their variants, on the other hand, are independent of magnetization and can also identify the edge of magnetic body. They not only excel at edge detection but can also be used to estimate the depth of the magnetic source producing the anomalies. In this study, we explore the utilization of the TDX derivative contour for estimating depth, assuming a vertical contact source.

**Theory**

TDX derivatives (Cooper and Cowan 2006) is basically a modified form of the tilt derivative (TDR), in which the total horizontal derivative (THDR) is normalized by the absolute value of the vertical derivative (VDR). This derivative, like other phase derivatives, has many interesting properties. The normalization makes this derivative unique, while the arctangent function restricts the output to between *0 and π/2*. Apart from the angular restriction and its independence from changes in magnetization, the derivative over the contact is much sharper with TDX than TDR (J. Derek Fairhead 2015)

# Depth to magnetic source estimation using TDX contour

as shown in Figure 1 over a two prism blocks model. (Cooper and Cowan 2006) described the TDX filter in the form of:

$$TDX = \tan^{-1}\left(\frac{THDR}{|VDR|}\right) = \tan^{-1}\left(\frac{\frac{dT}{dh}}{\left|\frac{dT}{dz}\right|}\right)$$

TDR and TDX derivatives share some interesting characteristics and can be related together.

$$TDR = \tan^{-1}\left(\frac{VDR}{THDR}\right), such\ |TDR| = \tan^{-1}\left(\frac{|VDR|}{THDR}\right)$$

$$TDX = \tan^{-1}(\tan^{-1}|TDR|)$$

Over contact or edge boundary, vertical derivative is approximately zero. So, |TDR| = 0 and TDX = π/2. This expression shows that the angle defined by TDX can only be positive since the absolute value of the vertical derivative is used to normalize the horizontal derivative. Thus, TDX is effectively π/2 - |TDR| which has value between 0° and 90° (J. Derek Fairhead 2015). This further shows that at the edge location, TDX has a maximum value since TDR (tilt) is zero at the same point. (Nabighian 1972) gives an expression for a vertical and horizontal derivative over a sloping contact model with horizontal location, *h*, and depth, *Z*, to the contact as:

$$\frac{dT}{dh} = 2kMc \sin d \frac{z\cos(2I - d - 90) + h\sin(2I - d - 90)}{h^2 + z^2}$$

$$\frac{dT}{dz} = 2kMc \sin d \frac{h\cos(2I - d - 90) - z\sin(2I - d - 90)}{h^2 + z^2}$$

Where *k* is the magnetic susceptibility contrast, *M* as the magnitude of the magnetic field, $c = 1 - cos^2i\ sin^2A$, *A* is the angle between the magnetic north and the horizontal *h-axis*, *i* is the ambient field inclination, $tan\ I = tan\ i\ /cos\ A$, and *d* is the dip. All trigonometric identities are in degrees. Assuming vertical contact and vertical magnetization (RTP). Substituting *d* as 90° and *A* as zero into the Nabighian expressions and TDX derivative above, TDX can be reduced to

$$TDX = \tan^{-1}\left(\frac{z}{h}\right) \dots\dots\dots\dots\dots\dots (1)$$

Equation 1 illustrates the correlation between the TDX amplitude and the depth of the magnetic contact. The maximum TDX amplitude is observed at h = 0 at the edge of the source. Additionally, by analyzing the contours, we can infer that the depth (z) is equivalent to the horizontal distance (h) when TDX is 45°. This demonstrates that TDX contours can be utilized to identify both the edge of the magnetic source (h = 0) at the 90° contour and the depth of contact-like structures (distance between 45° and 90° contours). Due to the inherent ambiguity of the 90° contour, depth estimation using the TDX map can be estimated by half the distance between the 45° contours on both sides of the 90° contour.

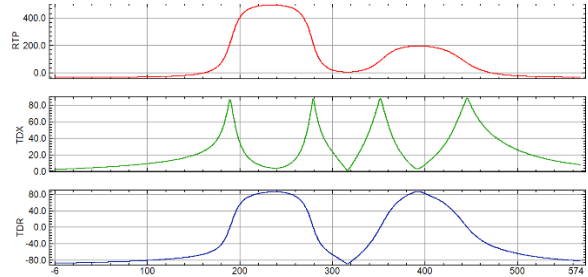

Figure 1: Comparison of TDX and TDR profile response over the two magnetic prim blocks in Figure 2. The vertical lines represent the contact or edge boundary. Maxima from the TDX and zero crossing from the TDR.

**Synthetic Example: Two Prism Blocks Model**

The methodology is applied to a synthetic model containing two vertical-sided prism blocks (Figure 2a). The depth to the top of the two prisms is known; 4km for the first prism, and 8km for the second prism. Both prisms with an infinite depth extent and a magnetization contrast of $10^{-4}$ A/m. The ambient magnetic field assumes vertical magnetization and has a declination of 0°. Figures 2b, c, and d show the total horizontal gradient, vertical gradient, and the absolute value of the VDR used to produce the TDX map (Figure 2e). Figure 2f only shows the TDX contour of interest. Observation of the 45° – 90° contours shows the depth to the top of the magnetic body can be approximated from the distance between two contours. While the distance between the two contours around the perimeter of the prism blocks is not uniform due to anomaly interference (Ahmed Salem, Williams, et al. 2007), the average distance between the contours for the first block is about 8km and 16km for the second prism block. The depth can be estimated by half the distance between the two contours. Figure 2g shows the estimated depth along the edge of the blocks.

**The Bishop Model**

Ever since the use of the 3D basement model was proposed by (S. Williams, Fairhead, and Flanagan 2002) to evaluate the effectiveness of depth to basement techniques, it has been used by researchers to accurately test techniques for estimating depth to magnetic source (For example: J. D. Fairhead, Williams, and Flanagan 2004; A. Reid, FitzGerald, and Flanagan 2005; Ahmed Salem et al. 2008; A. Salem et al. 2012; Ahmed Salem, Smith, et al. 2007;

# Depth to magnetic source estimation using TDX contour

Williams, Fairhead, and Flanagan 2005). This is because the model provides a realistic 3D test of the subsurface with a reasonable level of complexity. The model was created from a real topographic dataset from a part, 10.5km by 10.5km, of the volcanic tablelands area north of Bishop in California, with the original elevation model scaled by a factor of 30 in all directions to create a basin-sized setting.

The topography surface datum was adjusted in the depth direction, resulting in the structure being positioned below the subsurface, with the shallowest point having a depth of a few hundred meters in the NW and the deepest point having a depth of around 10 km in the SE (Figure 3f). Geologically speaking, numerous exposed en echelon fault scraps of

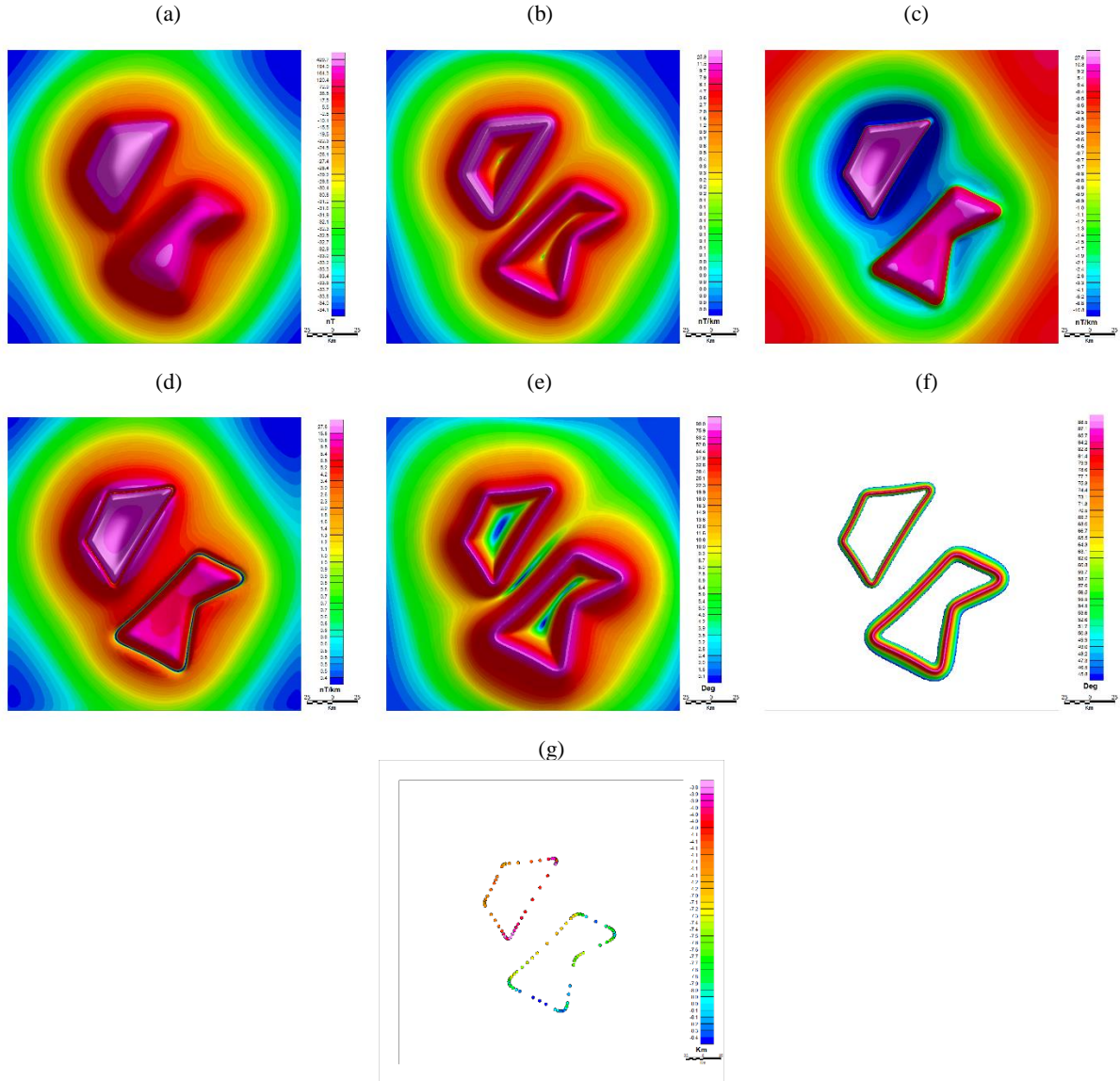

Figure 2: Two prism Model (a) Total Magnetic Intensity, (b) Total Horizontal Derivative, (c) Vertical Derivative, (d) Absolute value of Vertical Derivative, |VDR|, (e) TDX map, (f) contour of interest, 45° ≤ TDX ≤ 90°, and (g) Estimated depth along the depth of the block.

# Depth to magnetic source estimation using TDX contour

varying sizes, shapes, and orientations are visible in the area, as well as the existence of two major fault structures in the E-W and N-S direction (J. D. Fairhead, Williams, and Flanagan 2004). This model thus provides the uniqueness and complexities needed to test and validate any interpretation or automated depth determination technique. Using the simple 3D bishop model with a constant susceptibility of 0.001, we applied the methodology to contact. The method is tested on a two-prism theoretical model and the well-known 3D bishop model with a single basement susceptibility. The results agree with the known depth, providing evidence of the reliability of the method despite the restrictive nature of the assumption, especially for the Bishop model, where there are numerous fault structures.

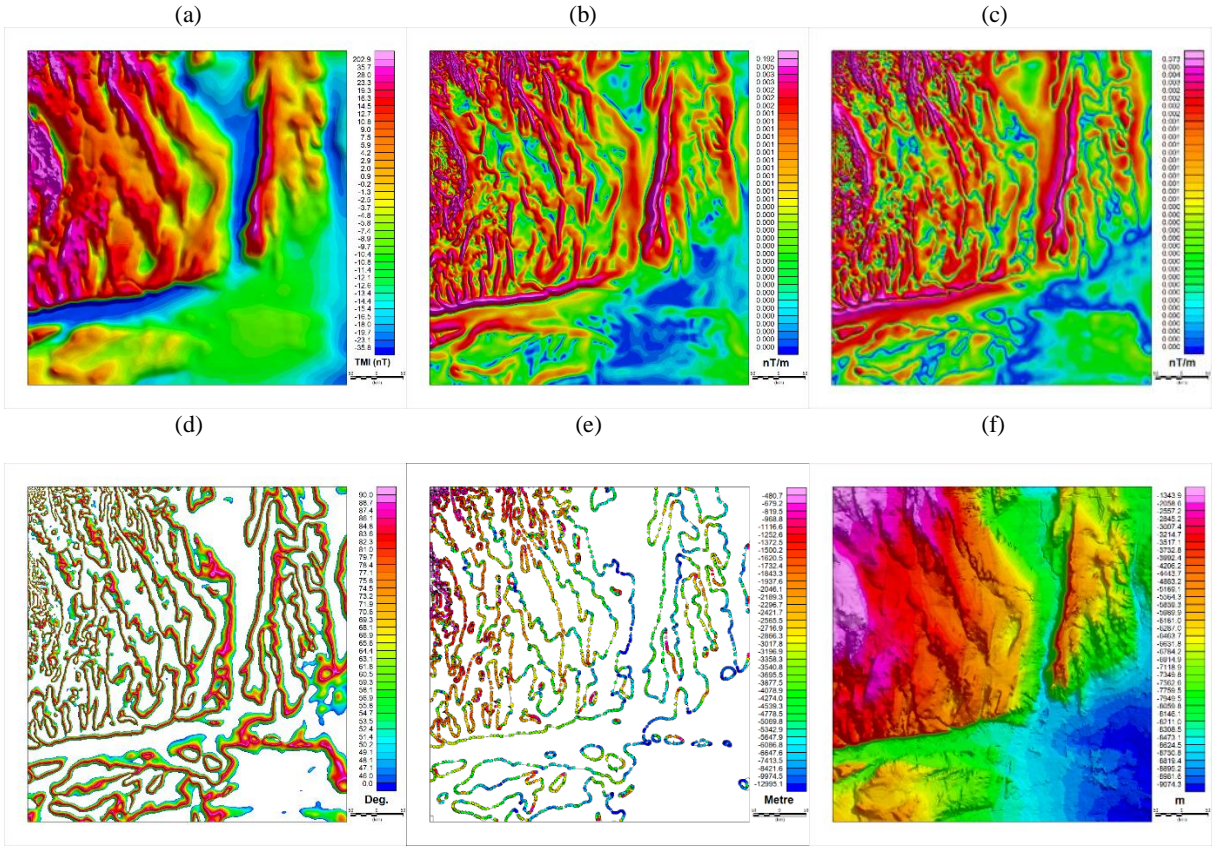

Figure 3: Bishop Models. (a) TMI response generated from the basement model, with uniform magnetic basement susceptibility, RTP'ed, and field strength of 50,000nT, (b) Total Gradient of the field, (c) Absolute value of the Vertical derivative, (d) TDX map between 45° and 90°, (e) Estimated depth along the edge of the block, and (f) Topography of the Model Magnetic Basement

estimate the depth of the basement using the contours of the TDX as explained above. By differentiating equation 1 above with respect to h, we can relate the total horizontal derivative of the TDX to the horizontal derivative of the tilt. As such, the depth along the edge of the body can be estimated. Figure 3e shows the estimated depth along the edge of the structure and 3f the magnetic basement model.

## Conclusions

This study presents the use of the contour of the TDX derivative as a depth estimation technique, assuming vertical

## Acknowledgments

The author would like to thank King Fahd University of Petroleum and Minerals for sponsoring this study.

# Depth to magnetic source estimation using TDX contour